\documentclass[reprint, aps, pra, floatfix, nofootinbib]{revtex4-1}

\usepackage[utf8]{inputenc}
\usepackage{amsmath,amssymb,amsfonts, amsthm, bbm, mathrsfs}
\usepackage{graphicx}
\usepackage{xcolor}

\DeclareMathAlphabet{\mathpzc}{OT1}{pzc}{m}{it}
\newcommand{\bra}[1]{\langle #1 \vert}
\newcommand{\ket}[1]{\vert #1 \rangle}

\allowdisplaybreaks[3]

\begin{document}

\title{Backaction-free measurement of quantum correlations\\via quantum time-domain interferometry}

\date{\today}

\author{Salvatore Castrignano}
\affiliation{Max-Planck-Institut f\"ur Kernphysik, Saupfercheckweg 1, 69117 Heidelberg, Germany}

\author{J\"org Evers}
\affiliation{Max-Planck-Institut f\"ur Kernphysik, Saupfercheckweg 1, 69117 Heidelberg, Germany}

\begin{abstract}
Time-domain interferometry (TDI) is a method to probe space-time correlations among particles in condensed matter systems. Applying TDI to quantum systems raises the general question, whether two-time correlations can be reliably measured without adverse impact of the measurement backaction onto the dynamics of the system. Here, we show that a recently developed quantum version of TDI (QTDI) indeed can access the full quantum-mechanical two-time correlations without backaction.  We further generalize QTDI to weak classical continuous-mode coherent input states, alleviating the need for single-photon input fields. Finally, we interpret our results by splitting the space-time correlations into two parts. While the  first one is  associated to  projective measurements and thus insensitive to backaction, we identify  the second contribution as arising from the coherence properties of the state of the probed target system, such that it is  perturbed or even destroyed by measurements on the system.
\end{abstract}

\maketitle

\section{Introduction}
Time-domain interferometry (TDI) is an experimental technique to probe space-time correlations among particles in condensed matter  systems~\cite{BARON,SMIRNOV,SMIRNOV2006,Saito2012,SAITO,Saito2017,Kaisermayr2001,1882-0786-2-2-026502,PhysRevLett.122.025301,doi:10.1063/1.4869541,doi:10.1063/1.5008868,Yamaguchi2018,PhysRevResearch.1.012008,Burkel_2000}. Such correlations are quantified by a generalization of the static pair distribution function~\cite{VANHOVE}, called dynamical couple correlation function (DCF), which is given by
\begin{equation}
\label{eqn: DefDCF}
G(\mathbf{r},t_1,t_2)\equiv \int_V  d^3 r' \: \langle \rho(\mathbf{r}',t_1)\rho(\mathbf{r}'+\mathbf{r},t_2) \rangle\,.
\end{equation}
Here, $\rho(\mathbf{r},t)$ is the density of particles at point $\mathbf{r}$ at time $t$ and $V$ is the volume occupied by the system. The DCF therefore describes correlations between having a particle at space-time point $(\mathbf{r}', t_1)$ {\it and} at $(\mathbf{r}' + \mathbf{r}, t_2)$, taking into account the entire sample via the integration over $\mathbf{r}'$. In the definition Eq.~(\ref{eqn: DefDCF}), the angular brackets stand for an average operation that can be either a classical ensemble average or a quantum mechanical expectation value, according to what description of the system is needed.

\begin{figure}[t]
   \centering
   \includegraphics[width=\columnwidth]{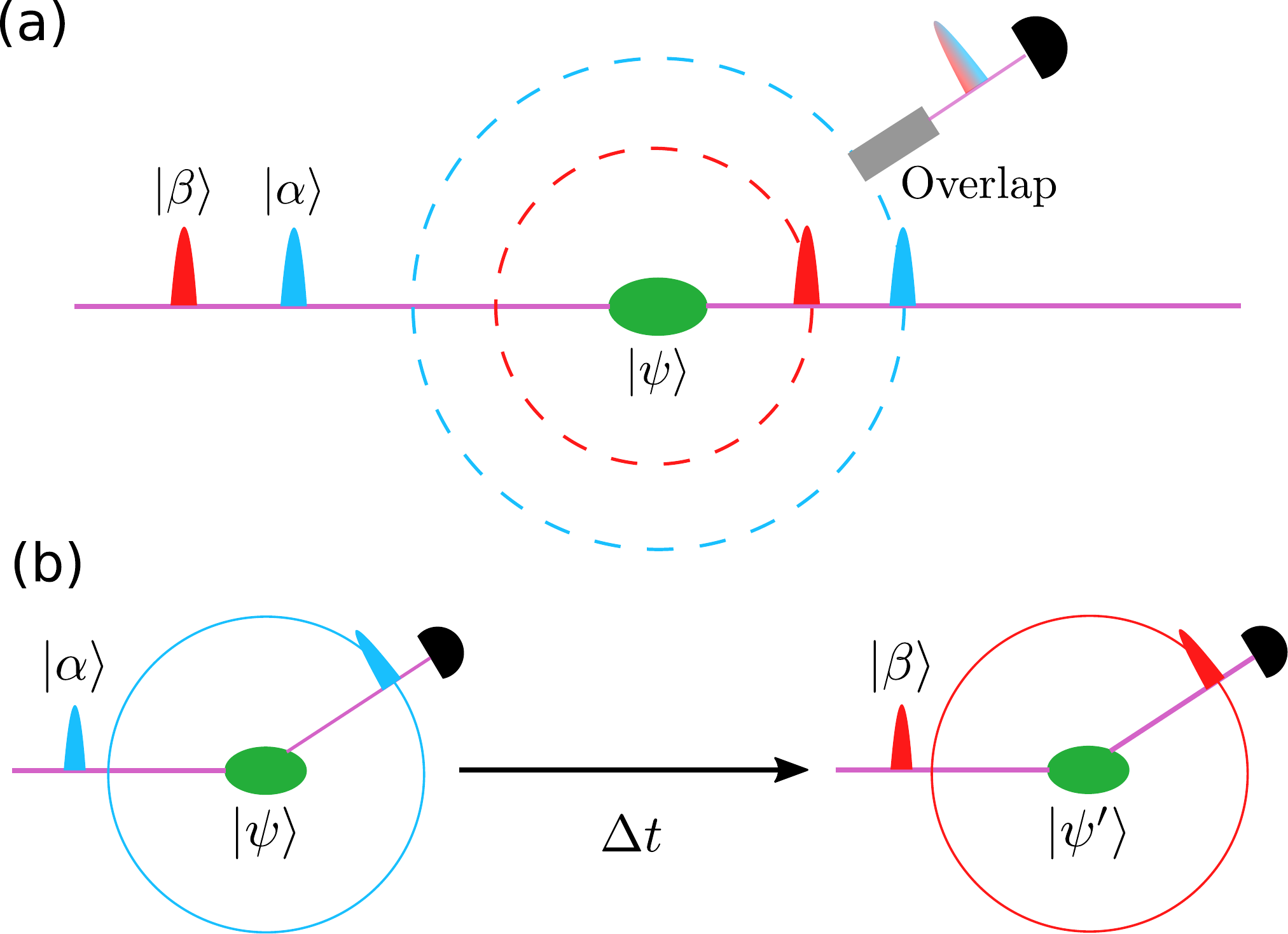}
   
   \caption{(Color online) Schematic illustration of quantum time-domain interferometry (QTDI). A distribution of particles in a target (green ellipse) is probed at two separate times via the scattering of two consecutive radiation pulses (denoted as blue $|\alpha\rangle$ and red $|\beta\rangle$). 
   (a) In case of a quantum mechanical target, the backaction of the first scattering event on the target could change its state such that the time evolution of the target until the second scattering event is modified. 
   In QTDI, only one of the two possible scattering events occurs in each repetition of the experiment, such that the measurement backaction does not affect the measurement outcome. Subsequently, an overlap stage erases the information about which of the two scattering possibilities took place,  in such a way that both contribute to the total scattered intensity. As a result, this intensity encodes the unperturbed quantum-mechanical space-time correlations among the particles in the target. 
   (b) In a classical variant of the scheme, the system is probed with two consecutive measurements in each experimental run. In this case, the first scattering event could change the state of the target, such that the time evolution until the second scattering is modified by the back-action on the target from the first measurement. }
  \label{fig: TDI}
\end{figure}

The basic idea of TDI is to scatter two x-ray pulses off of the system of interest at two different instants in time, see Fig.~\ref{fig: TDI}~\cite{BARON}. The  electric field scattered from each pulse then is proportional to the structure factor of the target evaluated at the respective scattering time. The two scattered fields are overlapped in time, and the intensity of the resulting field is measured by a detector placed far from the scattering zone. It can be shown that the average value of the recorded intensity then depends on the spatial Fourier transform of the DCF, known as the Intermediate Scattering Function (ISF)~\cite{BARON,SMIRNOV,SMIRNOV2006,Saito2012,Saito2017,Kaisermayr2001,1882-0786-2-2-026502,SAITO,PhysRevLett.122.025301,doi:10.1063/1.4869541,doi:10.1063/1.5008868,Yamaguchi2018,PhysRevResearch.1.012008,Burkel_2000}
\begin{equation}
\label{eqn: DefISF}
S(\mathbf{p},t_1,t_2)\equiv \int_V d^3 r \:G(\mathbf{r},t_1,t_2)e^{i\mathbf{p}\cdot \mathbf{r}}\,.
\end{equation}

In the experimental realizations of TDI so far, the two incoming pulses are generated by letting a synchrotron pulse interact with a moving metallic foil enriched with M\"ossbauer nuclei. As a result of this interaction, two forward-propagating pulses are produced: an instantaneous one, which is a copy of the original synchrotron pulse, and a trailing one with a duration given by the lifetime of the M\"ossbauer transition. A second foil, with the same characteristics as the first, but fixed, is put downstream of the target in order to re-overlap the scattered wavepackets. 
Depending on the chosen enriching M\"ossbauer nucleus, these realizations of TDI can probe correlations between particles on several space- and time-scales, ranging between $10^{-2}\div 1$~nm and $10^{-2} \div 10^4$~ns respectively~\cite{SAITO}. These scales make TDI a good candidate technique for closing a ``temporal gap'' in the investigation of microscopic dynamics in complex materials \cite{petra4}. With different methods to implement the incident double pulse and the overlap unit, even more general spatial- and temporal scales could become accessible, also beyond the x-ray regime. For example, a split-and-delay line may convert a single incident pulse into  two separate ones with a mutual delay tunable by the path length of the delay line. Another option is to consider two successive pulses from a frequency comb, a tool that is nowadays available in a broad range of frequencies, even with perspective of extension to the hard x-ray region~\cite{RevModPhys.87.637,Benko2014,Cavaletto2014,PhysRevSTAB.18.030711,2019arXiv190309317A}.

The TDI technique has been successfully used so far to study classical dynamics, such as slow diffusion in glass-forming fluids \cite{SAITO,Saito2012bis,Saito2014,doi:10.1063/1.4869541}, liquid crystals~\cite{doi:10.1143/JPSJ.81.023001,PhysRevResearch.1.012008}, viscous ionic liquids~\cite{C5CP02335A}, liquids with mesoscopic structures~\cite{Yamaguchi2018} and in ordered alloys \cite{Kaisermayr2001}. Recently, TDI has been proposed to probe correlations in strongly correlated materials \cite{SWISSFEL2017}, in which the interplay among the different degrees of freedom can produce correlations over many different space and time scales \cite{Dagotto257}. Understanding the behavior of this class of systems though calls for a quantum description of matter \cite{Dagotto257,QUINTANILLA,Keimer}.

However, the perspective of applying TDI to  quantum systems rises the general question whether two-time correlations such as Eq.~(\ref{eqn: DefDCF}) can faithfully be  measured. It is known indeed that the dynamics of a quantum system can be profoundly altered by the interaction with measurement devices \cite{Braginsky}. Because of this fact, in general quantum-mechanical time-correlations between two observables cannot be obtained simply by probing the observables consecutively, as illustrated in Fig.~\ref{fig: TDI}(b) for a scattering setting as in TDI. The back-action of the first measurement on the system affects the ongoing dynamics and in turn also the outcome of the second measurement. Therefore, the result obtained by correlating the outcomes of these two measurements  is modified by the external intervention on the system and, for this reason, does  not correspond to the time-correlations that would develop in an isolated system. This problem in general is present irrespective of how strong the back-action is, that is either if the measurements are direct and projective or weak and indirect, realized via the coupling with an auxiliary quantum system~\cite{Shimizu,Oehri}.

Therefore, naive consecutive measurements of the system are not a good option for measuring time correlations in the quantum realm and one has to rely on more involved measurement protocols. Recently, some protocols have been proposed for accomplishing this aim in the special case of space-time correlations between spin observables in spin lattice systems \cite{Knap,PhysRevA.96.022127}. In particular, in \cite{PhysRevA.96.022127} it was shown that time correlation functions between generic observables can be given as a sum of two terms. It was also found that the first of the two terms  can be measured by direct projective measurements of the correlated observables, therefore being insensitive to measurement back-action. Interestingly, in the case of spin observables, both terms can be separately accessed by coupling the spin lattice of interest to an auxiliary spin and by making direct projective measurements on both. By preparing the auxiliary spin in definite states, one or the other part of the above mentioned spin-spin time correlations can be obtained.

In a purely classical picture of TDI, one might conclude that this technique is not suitable for probing space-time correlations of particles in quantum systems because it is based on two consecutive interventions on the system [the scattering of the two radiation pulses, see Figure~\ref{fig: TDI}(b)]. In particular, the  presence of the classical field modifies the system before the second scattering takes place. 
However, recently we showed that this interpretation needs revision  in a full quantum theoretical analysis of TDI (QTDI), in which both the system and the incoming pulses are treated quantum mechanically~\cite{PhysRevLett.122.025301}. If the incident field comprises a single x-ray photon, then this photon either probes the particle density of the target at the earlier or at the later time, but it does not interact at both times. Therefore, in each repetition of the experiment, the system's dynamics is not modified by the probing field before the single interaction takes place. As a result, it was shown in ~\cite{PhysRevLett.122.025301} that QTDI provides access to the unperturbed ISF of the quantum system.

Here, we continue this development, and first show that QTDI does not require single x-ray photons as incident field, but may also operate with weak classical input fields. For this, we extend the QTDI analysis to continuous-mode coherent input states, and demonstrate that in leading order of a perturbative treatment, the photo-detection signal depends again on the full ISF of the quantum target. Next, we revisit the general splitting of the quantum-mechanical time-correlation functions found in \cite{PhysRevA.96.022127}, and clarify the meaning of the second term, explicitly showing that it is linked to the coherence properties of the state of the quantum system. This clearly shows that it is altered (if not destroyed) by any intervention on the system. We then specialize the general split form to the DCF and show that QTDI indeed is sensitive not only to the projective, but also to the coherent part of the DCF. Finally, we discuss how to measure  the full quantum-mechanical ISF and DCF using QTDI, evading the measurement backaction.

\section{Quantum Time-domain interferometry with coherent states of radiation}
\label{sec: TDI with coherent states}

In~\cite{PhysRevLett.122.025301}, we analyzed the most simple version of QTDI, with a single x-ray photon as incident state. In the first step of the QTDI scheme, the single incident photon is split in a time-bin entangled state~\cite{PhysRevLett.88.070402,PhysRevLett.103.017401}, i.e., in a state of the form 
\begin{align}\label{single-photon-entanglement}
 |1\rangle_{t}|0\rangle_{t+\tau} + e^{i\phi}|0\rangle_{t}|1\rangle_{t+\tau}\,,
\end{align}
in which it is not known whether the photon wavepacket arrives at the sample at a time $t$ or at a later time $t + \tau$. This state can be interpreted as the temporal variant of the photon behind a spatial double-slit, where the unknown which-way information now affects the position in time along a common trajectory. In addition to the original classical TDI scheme, we further proposed to control the relative phase $\phi$ in order to gain access to complex-valued quantities using this input state of light.
Like in classical TDI, one possible way of implementing the field in Eq.~(\ref{single-photon-entanglement}) is to let the single incident photon interact with a foil containing nuclei with narrow  M\"ossbauer resonances. This interaction separates the photon in a part which is delayed by $\tau$ due to the interaction with the nuclei, and one part which does not interact and thus is not delayed. In this case, the delay $\tau$ is random for each incident x-ray photon as it depends on the time of emission from the excited nuclei, which however does not impair the operation of the scheme. In this implementation, the relative phase $\phi$ of the two parts can be controlled using sudden motions of the nuclear target, as demonstrated, e.g., in~\cite{Helistoe1991,Schindelmann2002,Heeg2017,coherentcontrol}. 
In the next step of the scheme, the photon in the time-bin entangled state interacts with a target, which does not have to contain resonant nuclei. The light is assumed to scatter quasi-elastically, such that the temporal structure of the incident pulse remains unchanged, whereas the spatial part of the scattered wavepacket contains information about the target at two times. 
In a third step, the scattered wavepacket is sent through a second foil containing M\"ossbauer resonances, like the first one. The effect of the second foil on the scattering part of the photonic state is to split it into three temporally separated parts.
In the leading [trailing] part the photon was delayed in neither [either] of the M\"ossbauer foils. The interesting is the middle part, in which the photon was delyaed either in the first or the second foil, but not in both. This contribution therefore contains information on both  possible scattering times, and thereby on two-time information on the target. Analyzing the scattered intensity due to this component as a function of the momentum transfer and the relative phase $\phi$, the desired quantum mechanical ISF can be measured~\cite{PhysRevLett.122.025301}.

However, at present there is no established way to generate  single x-ray photons (in the quantum mechanical sense of Eq.~(\ref{single-photon-entanglement}), i.e., perfect anti-bunching) at sufficient rate. One approach to alleviate this limitation is to use post-selection. For this,  the x-rays incident onto the TDI setup can be monochromatized to the spectrum of the M\"ossbauer nuclei, such that no off-resonant photons perturb the sample to be probed. Then, in principle the detection events can be post-selected based on the number of detected photons, to {\it a posteriori} determine the measurement runs in which a certain number of photons was present in the setup. This, however, requires the observation of all possible scattering channels, which is challenging in practice. 

As an alternative route, in the following, we extend the theoretical analysis of QTDI to incoming continuous-mode coherent states, as a model for a classical incident field. We first use these states to model the incoming double pulses  and then proceed with the perturbative calculation of the state of target and radiation after the scattering. We then use this evolved state to evaluate the expected value of the photodetection signal. As a result, we show that this signal contains the desired information on the ISF of the quantum mechanical target as well. Finally we show how the relative phase of the incoming pulses can be exploited to measure the ISF.

\subsection{Incident radiation}

As a first step, we define the state of the incoming radiation via suitable  photon-wavepacket creation and destruction operators. Our aim is to characterize an initial state comprising two temporally separate wavepackets, each of which with a finite duration short compared to the time-scale of the internal dynamics of the target, propagating well-collimated along the $z$-direction, with a finite transverse area $\mathcal{A}$. In the temporal domain, the wavepacket can be written as,
\begin{equation}
\label{eqn: TimeWavepack}
 \alpha(z-ct)e^{i (k_0 z - \omega_0 t + \phi_\alpha)}\,,
\end{equation}
where $\omega_0=k_0/c$ center frequency with wave number $k_0$ and $\phi_\alpha$ the overall phase of the wavepacket. Via the Fourier transform, the corresponding frequency space wavepacket is obtained as a superposition of modes having wavevectors parallel to the $z$ axis with amplitude $\tilde{\alpha}(k_z)$. 
Using this, we define the creation operator of the corresponding photon-wavepacket as
\begin{equation}
\label{eqn: CreWavePack}
a_\alpha ^\dagger  \equiv \frac{1}{\sqrt{\mathcal{A}}} \int_{-\infty} ^\infty dk_z \: \tilde{\alpha}(k_z) \: a^\dagger _{k_z}\,,
\end{equation}
and the corresponding destruction operator $a_\alpha$ as the hermitian conjugate of (\ref{eqn: CreWavePack}).
By means of these operators, we furthermore define the displacement operator
\begin{equation}
\label{eqn: DefDisplOp}
D(\alpha)\equiv e^{(a^\dagger _\alpha - a _\alpha)}\,.
\end{equation}
Acting with this operator on the vacuum state $\ket{0}$, a continuous-mode coherent state is obtained \cite{loudon2000quantum}.

To model two  consecutive but spatially separated wavepackets, we analogously define a second set of creation, destruction and displacement operators corresponding to a second photon-wavepacket. We call the distribution of amplitudes of this second wavepacket $\tilde{\beta}(k_z)$ and assume that it is also peaked around $k_0$ and that its envelope $\beta(z-ct)$ doesn't overlap with $\alpha(z-ct)$. Moreover, we take its initial phase $\phi_\beta$ to be different from $\phi_\alpha$. The second set of operators is
\begin{align}
a^\dagger _\beta &\equiv \frac{1}{\sqrt{\mathcal{A}}} \int_{-\infty} ^\infty dk_z \, \tilde{\beta}(k_z) a^\dagger _{k_z} \,,\\
D(\beta) &\equiv e^{(a^\dagger _\beta - a _\beta)}\,.
\end{align}
Note that for simplicity the two wavepackets are assumed to have the same transverse area $\mathcal{A}$, without loss of generality. 

With these definitions at hand, we finally define the state of the incoming radiation as
\begin{equation}
\label{eqn: ContModeCohState}
\ket{\alpha,\beta}\equiv D(\alpha) D(\beta) \ket{0}\,.
\end{equation}
Eq.~(\ref{eqn: ContModeCohState}) represents again a classical-like state, in which the space-time dependency of the radiation is given by the superposition of the  $\alpha$ and $\beta$ wavepackets. One way of generating this state involves a split-and-delay line, which may convert a single coherent state wavepacket into two separate ones with a mutual delay tunable by the path length of the delay line. Another option is to consider two successive pulses from a frequency comb, a tool that is nowadays available in a range of frequencies that goes from the infrared to soft x-ray with perspective of extension to the hard x-ray region~\cite{PhysRevSTAB.18.030711,2019arXiv190309317A}.

\subsection{Scattering on the target}

Next, we calculate the scattering of the incident state Eq.~(\ref{eqn: ContModeCohState}) off of the target system.  
We assume that at initial time $t_0=0$ the incident radiation has not yet reached the target, such that we can make a product ansatz
\begin{equation}
\label{eqn: InitState}
\ket{\Psi_0}=\ket{\psi}\ket{\alpha,\beta}\,,
\end{equation}
with $\ket{\psi}$ the state of the target at $t_0=0$. In the following, we assume that the intensity of the incoming radiation is low enough to meaningfully compute the evolution of state (\ref{eqn: InitState}) by a first-order perturbative calculation. As we are interested in the scattering of radiation by the spatial structure of the target and consider wavepackets whose spectra do not resonate with the internal energy level structure of the target itself, the term of the matter-radiation interaction Hamiltonian that dictates the dynamics of the composite target-radiation system is the diamagnetic term \cite{HAURIEGE}, that explicitly reads
\begin{align}
\label{eqn: IntHam}
H_I=&r_e \frac{\hbar c^2}{4 \pi} \int d^3 k \int d^3 k' \int_V d^3r \frac{1}{\sqrt{\omega_k \omega_{k'}}} \nonumber\\
& \times a^{\dagger}_{\mathbf{k}} \: a_{\mathbf{k}'}\: \rho(\mathbf{r})\: e^{-i (\mathbf{k}-\mathbf{k}') \cdot \mathbf{r}}\,,
\end{align}
where $r_e$ is the classical radius of the electron.

The explicit first-order calculation (see Appendix~\ref{app-perturb} for details) shows that the evolved state of the composite system at a time $t$ after the incident radiation crossed the target is given by the sum of three contributions,
\begin{equation}
\label{eqn: FinalStateImplicit}
\ket{\Psi(t)} \simeq \ket{\psi}\ket{\alpha,\beta}+\ket{\delta \Psi_\alpha (t)}+\ket{\delta \Psi_\beta (t)}\,.
\end{equation}
Under the assumption that the internal dynamics of the target is slow as compared to the crossing time of each of the wavepackets through the target, the explicit forms of the latter two parts can be evaluated to give
\begin{align}
\label{eqn: ExplDeltaAleph}
\ket{\delta \Psi_\alpha (t)} =&
i  \frac{r_e c}{4 \pi \omega_0 \sqrt{\mathcal{A}} }  \int_V d^3r \:\rho(\mathbf{r},t_\alpha) \:\ket{\psi} \nonumber \\
&  \times \int d^3 k \: \tilde{\alpha}  (|\mathbf{k}|) \:e^{i \omega_k \frac{z}{c}}\:e^{-i \mathbf{k}\cdot\mathbf{r}}
 \: a^{\dagger}_{\mathbf{k}}\: \ket{\alpha,\beta}\,,\\[2ex]
\label{eqn: ExplDeltaGimel}
\ket{\delta \Psi_\beta (t)} =&
i  \frac{ r_e c}{4 \pi \omega_0 \sqrt{\mathcal{A}}}   \int_V d^3r \:\rho(\mathbf{r},t_\beta)\: \ket{\psi} \nonumber \\
& \int d^3 k \, \tilde{\beta}   (|\mathbf{k}|) \:e^{i \omega_k \frac{z}{c}} \:e^{-i \mathbf{k}\cdot\mathbf{r}}
 \: a^{\dagger}_{\mathbf{k}} \:\ket{\alpha,\beta}\,,
\end{align}
where $t_\alpha$ and $t_\beta$ are the times at which the wavepackets $|\alpha\rangle$ and $|\beta\rangle$ interacted with the target, respectively. 

The three terms in Eq.~(\ref{eqn: FinalStateImplicit}) can be interpreted in a straightforward way, see Figure~\ref{fig: TDI}. The first term is the zeroth order contribution and represents the case in which the light passed through the target without interaction. 
The two other contributions characterize a scattering of either $|\alpha\rangle$ or $|\beta\rangle$ on the target.  During the interaction with the target, either a photon from the incoming radiation is destroyed by the target at time $t_\alpha$ and created in a spherically symmetric wavepacket with mode amplitudes $\tilde{\alpha}(|\mathbf{k}|)$, or it is destroyed at $t_\beta$ and created in a spherically symmetric wavepacket with mode amplitudes $\tilde{\beta}(|\mathbf{k}|)$. The final state thus contains a single-photon spherical wavepacket which carries information about the target at two different times, even though the radiation interacted only once with the target, as it is apparent from the first-order perturbative treatment. Therefore, even though the initial state is a classical-like state, due to the single-scattering approximation a similar final state is encountered as in our previous work~\cite{PhysRevLett.122.025301} in which we assumed an incoming single-photon state. As a result, we will again see in the following that by making the two pathways indistinguishable, the full ISF can be retrieved via the detection of the scattered photon.

\subsection{Scattered light intensity}
\label{sec: The recorded intensity}

Next, we consider the detection of the scattered light characterized by Eq.~(\ref{eqn: FinalStateImplicit}). The signal produced at time $t$ by a photo-detector placed at a point $\mathbf{R}$  is given by~\cite{loudon2000quantum}
\begin{align}
\label{eqn: ScattRate}
I(\mathbf{R},t) =\bra{\Psi(t)}  E^{(-)}(\mathbf{R},t)E^{(+)}(\mathbf{R},t) \ket{\Psi(t)}\nonumber \\
=\sum_{l,h=\alpha,\beta} \bra{\delta \Psi_l(t)} E^{(-)}(\mathbf{R},t)E^{(+)}(\mathbf{R},t)  \ket{\delta \Psi_h(t)}\,,
\end{align}
where in the second step we have assumed that the detector is placed outside the incident radiation such that the zeroth order contribution to $\ket{\Psi(t)}$ can be neglected. This condition can be implemented in the calculation of the intensity Eq.~(\ref{eqn: ScattRate}) by substituting the term $a^\dagger_{\mathbf{k}}\, \ket{\alpha,\beta}$ appearing in Eqs.~(\ref{eqn: ExplDeltaAleph}) and (\ref{eqn: ExplDeltaGimel}) with $a^\dagger_{\mathbf{k}} \,\ket{0}$ (details on the  calculation of the detection signal are given in Appendices~\ref{app-signal} and~\ref{app-equiv}). After this substitution, the calculation of the intensity can be simplified by introducing the projector on the vacuum state of the electromagnetic field in between the two electric field operators, 
\begin{align}
\label{eqn: ScattRateAmpl}
I(\mathbf{R},t) = \sum_{l,h=\alpha,\beta} &\bra{\delta \Psi_l(t)} E^{(-)}(\mathbf{R},t)\ket{0} \nonumber \\
&\times \bra{0} E^{(+)}(\mathbf{R},t)  \ket{\delta \Psi_h(t)}\,.
\end{align}
The quantity $\bra{0} E^{(+)}(\mathbf{R},t)  \ket{\delta \Psi_l(t)}$ ($l=\alpha,\beta$) then represents the probability amplitude that a photon from the $l$-th incoming wavepacket is absorbed by the detector after being scattered from the target \cite{SCULLY}, and the right-hand side of Eq.~(\ref{eqn: ScattRateAmpl}) is just the squared modulus of the sum of these amplitudes for the two possible scattering channels for a photon. 
The explicit form of the detection amplitude for channel $l\in \{\alpha,\beta\}$ is (see Appendix~\ref{app-signal} for details)
\begin{align}
\label{eqn: DetAmpl}
&\bra{0}E^{(+)}(\mathbf{R},t)\ket{\delta \Psi_l (t)}  \nonumber \\
&\quad=\frac{r_e}{2}\sqrt{\frac{\hbar \omega_0}{2(2\pi)^3 \epsilon_0 \mathscr{A}}}\frac{ e^{i(k_0 R -\omega_0 t)}}{R}\: e^{i\phi_l}\,l(R-ct) \nonumber \\
&\qquad \times \int_V d^3r \, \rho(\mathbf{r},t_l)\ket{\psi}e^{-i \mathbf{p}\cdot\mathbf{r}}\,.
\end{align}
%
Substituting this expression and its conjugate into Eq.~(\ref{eqn: ScattRateAmpl}), we obtain the final form of the intensity
\begin{align}
\label{eqn: DetectRateFinal}
I(\mathbf{R},t) &= I_0 \Big\{ |\alpha_{R,t}|^2 \:S\big(\mathbf{p},t_\alpha   ,t_\alpha \big) +|\beta_{R,t}|^2\: S\big(\mathbf{p},t_\beta,t_\beta \big)\nonumber \\
&+ 2 \, \text{Re} \Big[ \alpha_{R,t}^*\, \beta_{R,t} \,e^{i(\phi_\beta-\phi_\alpha)} S\big(\mathbf{p},t_\alpha,t_\beta) \Big] \Big \}\,,
\end{align}
with
\begin{subequations}
\begin{align}
 I_0 &= \frac{r^2_e}{\mathcal{A}} \frac{\hbar \omega_0}{2  \epsilon_0 (2\pi)^3}\frac{1}{ R^2}\,,\\
 \alpha_{R,t} &=\alpha(R-ct)\,,\\
 \beta_{R,t} &= \beta(R-ct)\,.
\end{align}
\end{subequations}
Further,  $\mathbf{p}=k_0(\mathbf{R}/R-\hat{\mathbf{z}})$ is the momentum  exchanged during the scattering.

The first two terms in the upper line of Eq.~(\ref{eqn: DetectRateFinal}) relate to the probability  that the photon is detected after it has been scattered from  either the pulse $\alpha$ at time $t_\alpha$ or from $\beta$ at time $t_\beta$, respectively. These terms therefore each only contain information on the target at a single instance in time, but not on the correlations between different times.
The third term instead arises from the interference between the probability amplitudes associated to these two channels and contains the  desired two-time ISF of the target. It depends on the spatio-temporal overlap of the two scattered wavepackets $\alpha_{R,t}$ and $\beta_{R,t}$. In order to enhance the contribution to the detection signal due to the ISF in Eq.~(\ref{eqn: DetectRateFinal}), the scattered wavepackets $\alpha_{R,t}$ and $\beta_{R,t}$ must be overlapped before they reach the detector, such that $\alpha_{R,t}^*\, \beta_{R,t}$ becomes non-zero. In the setup here considered this can be accomplished, e.g., using a delay line similar to that which can be used to create the incident field with two pulses separated in time.

If the two wavepackets do not overlap, then the time of arrival of the signal at the detector reveals whether the scattering occurred at $t_\alpha$ or at $t_\beta$. As a result, one of the amplitudes $\alpha_{R,t}$ or $\beta_{R,t}$ is zero, and the interference term vanishes. This can be understood in analogy to a double-slit experiment. If the path is know by which the photon travels to the detector, no interference occurs. In contrast, for overlapping scattered wavepackets, the time of arrival does not define the actual scattering path, and thereby essentially removes the ``which path information''. Then, the interference term contributes. 
To enhance the interference contribution, $\alpha_{R,t}$ and $\beta_{R,t}$ should ideally overlap in space and time, and further have the same shape. The latter is automatically fulfilled if the two incident wavepackets have equal shape, since their envelopes are indeed preserved by the scattering on the target, see Eq.~(\ref{eqn: DetAmpl}). Double-pulses with two identically shaped wavepackets can be realized, e.g., by generating the double-pulse from a single pulse via a split-and-delay line.   In the original realization of TDI with M\"ossbauer foils, the overlap of the scattered wavepackets is achieved by passing the scattered light through a second foil, which again absorbs and re-emits part of the scattered light. If the two foils in the setup are equal, also the envelopes of the two interfering wavepackets are equal.

Assuming equal shapes of the two interfering wavepackets, the envelopes appearing in Eq.~(\ref{eqn: DetectRateFinal}) factorize and this expression simplifies to
\begin{align}
\label{eqn: DetectRateFinalEqualAmpl}
I(\mathbf{R},t) &= \bar{I}_0(\mathbf{R},t) \Big\{  \:S\big(\mathbf{p},t_\alpha   ,t_\alpha \big) + \: S\big(\mathbf{p},t_\beta,t_\beta \big)\nonumber \\
&+ 2 \, \text{Re} \Big[ \,e^{i(\phi_\beta-\phi_\alpha)} S\big(\mathbf{p},t_\alpha,t_\beta) \Big] \Big \}\,,
\end{align}
with prefactor  $\bar{I}_0(\mathbf{R},t) = I_0 |\alpha_{R,t}|^2 = I_0 |\beta_{R,t}|^2$.

\subsection{\label{sec:recovery}Recovery of the ISF}

With equation Eq.~(\ref{eqn: DetectRateFinalEqualAmpl}) at hand, we can finally discuss how to extract the desired ISF from the scattered intensity. For this, we exploit its dependence on the relative phase $\phi\equiv \phi_\beta - \phi_\beta$ and rewrite
\begin{align}
\label{eqn: Interferogram}
I(\phi) = \bar{I}_0(\mathbf{R},t)  \bigg\{ S(\mathbf{p},t_\alpha,t_\alpha)+S(\mathbf{p},t_\beta,t_\beta) +\nonumber \\
+ 2 |S(\mathbf{p},t_\alpha,t_\beta)| \, \cos \big[ \phi+\text{arg}\, S(\mathbf{p},t_\alpha,t_\beta) \big] \bigg\}\,,
\end{align}
where the polar representation 
\begin{align}
S(\mathbf{p},t_\alpha,t_\beta)\equiv |S(\mathbf{p},t_\alpha,t_\beta)|\:e^{i \, \text{arg}\,S(\mathbf{p},t_\alpha,t_\beta)} 
\end{align}
for the ISF has been adopted.
The experiment then is repeated for  different values of the relative phase $\phi\equiv \phi_\beta - \phi_\beta$, but fixed values of the exchanged momentum $\mathbf{p}$. From Eq.~(\ref{eqn: Interferogram}) we find that the data has a cosine-dependence on $\phi$, which can be extracted by fitting a model $ A + B \cos(\phi + \phi_0)$ to the data. $B$ then is proportional to the absolute value $|S(\mathbf{p},t_\alpha,t_\beta)|$ of the ISF, and $\phi_0$ determines its phase $\text{arg}\,S(\mathbf{p},t_\alpha,t_\beta)$.

Note that as discussed in Section~\ref{sec:backaction-tdi}, it is not necessary to recover the full ISF if the main goal is to verify the presence of quantum correlations in the target.

\section{Measurement backaction in Quantum Dynamical Correlation Functions}
\label{sec: Quantum Dynamical Correlation Functions}

In this Section, we study the effect of measurement backactions in the measurement of two-time correlation functions. In particular, we split the full correlation function into a projective part which is the contribution due to projective measurements, and  a remaining ``coherent'' part which contains the remaining contributions that are lost in a projective measurement scheme. We then apply this analysis to the dynamical couple correlation function (DCF). Finally, we show that TDI enables one to access both, the projective and the coherent part of the DCF, and thereby allows for a backaction-free measurement of the full DCF. 

\subsection{General case}

We start by considering a generic quantum system, which we assume to be in the initial state $\ket{\psi}$ at time $t_0=0$, and which we probe with two observables $A$ and $B$ at times $t_1 \leq t_2$, respectively. We further denote the set of eigenstates of operator $A$ [$B$] as $\{|a_j\rangle\}$ [$\{|b_l\rangle\}$], with~\footnote{Here, we for simplicity assume the notation for a discrete eigenvalue spectrum.}
\begin{subequations}
\begin{align}
 A|a_j\rangle &= a_j|a_j\rangle\,,\\
 B|b_l\rangle &= b_l|b_l\rangle\,.
\end{align}
\end{subequations}
The correlation function for the two operators is defined as
\begin{equation}
\label{eqn: DefCorr}
C(t_1,t_2) \equiv \bra{\psi} A(t_1) B(t_2) \ket{\psi}\,.
\end{equation}
Here, $A(t_1)$ and $B(t_2)$ are the two operators in the Heisenberg picture, given by 
\begin{subequations}
\begin{align}
\label{eqn: HeisenOp}
A(t_1)&\equiv U(t_1)^{\dagger} A U(t_1)\,,\\
B(t_2)&\equiv U(t_2)^{\dagger} B U(t_2)\,,
\end{align}
\end{subequations}
via the time evolution operator of the system $U(t)$.
Evaluating the complex conjugate of Eq.~(\ref{eqn: DefCorr}),
\begin{equation}
\label{eq: conjugate}
C(t_1,t_2)^* = \bra{\psi}  B(t_2)A(t_1) \ket{\psi}\,,
\end{equation}
we find that $C(t_1,t_2)$ is in general a complex-valued function, if the two observables do not commute, which clearly distinguishes the quantum correlation function from its classical counterpart.

In order to split Eq.~(\ref{eqn: DefCorr}) into a projective and a remaining coherent part introduced, we decompose $A(t_1)$ and $B(t_2)$ into their spectral representations, that is
\begin{subequations}
\label{eqn: ADecomp}
\begin{align}
A(t_1)&=\sum_j a_j \Pi_{a_j} (t_1)\,, \\
B(t_2)&=\sum_l b_l \Pi_{b_l} (t_2)\,,
\end{align}
\end{subequations}
where $\Pi_{a_j}(t_1)$ is the projector onto state $|a_j\rangle$ in the Heisenberg picture at time $t_1$, and $\Pi_{b_l}(t_2)$ the corresponding projector onto $|b_l\rangle$. Further, we use the completeness relation
\begin{align}
\label{eq: complete}
1=\sum_{m} \Pi_{a_{m}}(t_1)\,.
\end{align}
Introducing Eqs.~(\ref{eqn: ADecomp}) into Eq.~(\ref{eqn: DefCorr}) and further inserting Eq.~(\ref{eq: complete}) to the right of $B(t_2)$, we obtain
\begin{align}
\label{eqn: CorrInc}
C(t_1,t_2)&=\sum_{j,m,l} a_j \, b_l\bra{\psi}\Pi_{a_j}(t_1)\Pi_{b_l}(t_2)\Pi_{a_{m}}(t_1)\ket{\psi} \nonumber \\
&=\mathscr{C}(t_1,t_2) + K(t_1,t_2)\,,\\[2ex]
\mathscr{C}(t_1,t_2) &= \sum_{j,l} a_j \, b_l\bra{\psi}\Pi_{a_j}(t_1)\Pi_{b_l}(t_2)\Pi_{a_j}(t_1)\ket{\psi}\,,\\
K(t_1,t_2) &= \sum_{j\neq m \atop  l} a_j \, b_l \bra{\psi}\Pi_{a_j}(t_1)\Pi_{b_l}(t_2)\Pi_{a_{m}}(t_1)\ket{\psi}\,.
\end{align}
To show that this is indeed the splitting we are searching for, we evaluate the action of the projectors $\Pi_{a_j}(t_1)$ on $\ket{\psi}$\,,
\begin{align}
\label{eqn: ExpCoeff}
\Pi_{a_j} (t_1)\ket{\psi} &= U(t_1)^\dagger \Pi_{a_j} U(t_1)\ket{\psi} \nonumber \\[2ex]
 &=U(t_1)^\dagger \Pi_{a_j} \sum_j c_j(t_1) \ket{a_j} \nonumber \\
 &=  c_j(t_1)U^{\dagger}(t_1)\ket{a_j}\,.
\end{align}
Here, $c_j(t_1)$ are defined via the decomposition of $U(t_1)\ket{\psi}$ into eigenstates of $A$,
\begin{align}
 U(t_1) |\Psi\rangle = \sum_j c_j(t_1) |a_j\rangle\,.
\end{align}
Substituting Eq.~(\ref{eqn: ExpCoeff}) into Eq.~(\ref{eqn: CorrInc}), we find
\begin{align}
\label{eqn: ProjCorr}
\mathscr{C}(t_1,t_2)&=\sum_{j,l} a_j b_{l} |c_j(t_1)|^2  \times \nonumber \\
  & \qquad \times \bra{a_j} U(t_1) \Pi_{b_l}(t_2) U(t_1)^\dagger\ket{a_j} \,,\\[2ex]
\label{eqn: CohCorr}
K(t_1,t_2)&=\sum_{j \neq m }\sum_l a_j b_{l}\, c_j(t_1)^* c_{m}(t_1)\times \nonumber \\
& \qquad \times   \bra{a_j} U(t_1) \Pi_{b_l}(t_2) U(t_1)^\dagger \ket{a_m}\,.
\end{align}
%
The term $\mathscr{C}(t_1,t_2)$ describes the correlation between two projective measurements: First, a measurement of $A$ at $t_1$, which projects the state into $|a_j\rangle$ with a probability of $|c_j(t_1)|^2$. The sum over $j$ accounts for all possible measurement outcomes. Subsequently, the state $|a_j\rangle$ evolves from $t_1$ to $t_2$, at which time a second measurement is applied, using  $B$. The possible outcomes of this second measurement are taken into account via the sum over $l$. In other words, the quantity $|c_j(t_1)|^2 \bra{a_j} U(t_1) \Pi_{b_{l}}(t_2) U(t_1)^\dagger\ket{a_j}$ in Eq.~(\ref{eqn: ProjCorr}) can be interpreted as the joint probability, calculated according to the Born rule, of obtaining the outcomes $a_j$ and $b_l$ from the two consecutive direct measurements of $A$ and $B$ at times $t_1$ and $t_2$. Therefore, from now on, we will denote $\mathscr{C}(t_1,t_2)$ as the {\it projective part of the quantum dynamical correlation function}. It resembles classical correlation functions, which are defined only by the joint statistics of  events.

The second contribution $K(t_1,t_2)$ in Eq.~(\ref{eqn: CohCorr}) depends instead on the coherences $c_j(t_1)^* c_m(t_1)$ ($j\neq m$) between the contributions of $|a_j\rangle$ and $|a_m\rangle$ of the state of the system at time $t_1$. As a result, this term immediately vanishes upon a projective measurement, which renders all but one of the coefficients $a_j(t_1)$ zero. We will therefore refer to this second contribution as the {\it coherent part of the dynamical correlation function}.

As shown in Eq.~(\ref{eq: conjugate}), the quantum mechanical correlation function $C(t_1,t_2)$ in general is a complex quantity, in contrast to the corresponding classical counterpart. From Eq.~(\ref{eqn: ProjCorr}), it directly follows that $\mathscr{C}(t_1,t_2)$ is a real quantity. In contrast, the real and imaginary parts of the coherent contribution $K(t_1,t_2)$ in general both are non-vanishing. As a result, we can attribute possible imaginary parts of  $C(t_1,t_2)$ to the coherent part of the correlation function. 
This is of practical relevance for experiments designed to measure dynamical correlations of quantum systems, because the detection of a non-vanishing imaginary part of the correlation function establishes the quantum mechanical nature of the probed correlations. It is important to note, however, that a purely real correlation function does not imply the classical nature of the correlations. On the one hand, the correlation function can be purely real, even though the system is in a quantum mechanical superposition of different eigenstates (an example is given in \citep{PhysRevLett.122.025301}). 
On the other hand, even in case of a vanishing coherent part, such that  the total dynamical correlation would be given by the projective part alone, it may not be appropriate to understand it as a classical correlation. The reason is that the dynamics of the system between the two successive projective measurements is governed by the quantum mechanical unitary evolution, so that the definition of the joint probability in Eq.~(\ref{eqn: ProjCorr}) is based on a quantum mechanical background and it is not granted that classical models of the dynamics can reproduce the same joint probability.

\subsection{Dynamical couple correlation function and intermediate scattering function}
\label{sec: The case of the dynamical couple correlation function}

Next, we apply the  previous general discussion to the specific case of interest in the present work, and show how the splitting into a projective and a coherent part applies to the DCF, and how this modifies the features of the ISF, for a quantum system composed of $N$ particles. To this end, we define the decomposition of the DCF into the projective and the coherent part as
\begin{align}
\label{eq: splitG}
 G (\mathbf{r},t_1,t_2) = \mathscr{G} (\mathbf{r},t_1,t_2) + \Gamma (\mathbf{r},t_1,t_2)\,. 
\end{align}

We start with definition  of the DCF in Eq.~(\ref{eqn: DefDCF}), where now the angular brackets are meant as the expectation value on the initial state of the system, $\ket{\psi}$, and the $\rho$ are particle density operators acting on the state of the $N$ particles composing the system. An eigenstate basis for these operators is given by the $N$-particle position eigenstates $\ket{\mathbf{r}_1,\ldots,\mathbf{r}_N}$. Thus, if the system is in one of these eigenstates, then the density of particles at position $\mathbf{r}$ will assume a definite eigenvalue that we will call $\rho(\mathbf{r};\mathbf{r}_1,\ldots,\mathbf{r}_N)$.

With these definitions, we can now apply Eqs.~(\ref{eqn: ProjCorr}) and (\ref{eqn: CohCorr}) to $G(\mathbf{r},t_1,t_2)$ to obtain the projective part
\begin{align}
\label{eqn: ProjDCF}
\mathscr{G} (\mathbf{r},t_1,t_2) &=\int_V d^3 r' \int_V d^3 r_{1\ldots N}  \int_V d^3 \tilde{r}_{1\ldots N} & \nonumber\\
&\times \rho(\mathbf{r}';\mathbf{r}_1,\ldots,\mathbf{r}_N) \:\rho(\mathbf{r}'+\mathbf{r};\tilde{\mathbf{r}}_1,\ldots,\tilde{\mathbf{r}}_N) \nonumber \\
&\times \big| \bra{\tilde{\mathbf{r}}_1,\ldots,\tilde{\mathbf{r}}_N} U(t_2-t_1) \ket{\mathbf{r_1},\ldots,\mathbf{r}_N}\big| ^2 \nonumber \\
& \times \big| \psi(\mathbf{r}_1,\ldots,\mathbf{r}_N,t_1) \big| ^2\,,
\end{align}
and the coherent part
\begin{align}
\label{eqn: CohDCF}
\Gamma (\mathbf{r},t_1,t_2) &=\int_V d^3 r'
\int_V d^3 r_{1\ldots N}  \int_V d\tilde{r}_{1\dots N} \int_{\overline{V}} d^3 \overline{r}_{1\dots N}\nonumber\\
&\times  \rho(\mathbf{r}';\mathbf{r}_1,\ldots,\mathbf{r}_N)\: \rho(\mathbf{r}'+\mathbf{r};\tilde{\mathbf{r}}_1,\ldots,\tilde{\mathbf{r}}_N) \nonumber\\
&\times\psi(\mathbf{r}_1,\ldots,\mathbf{r}_N,t_1)^* \: \psi(\overline{\mathbf{r}}_1,\ldots,\overline{\mathbf{r}}_N,t_1) \nonumber\\
&\times\bra{\tilde{\mathbf{r}}_1,\ldots,\tilde{\mathbf{r}}_N} U(t_2-t_1) \ket{\mathbf{r}_1,\ldots,\mathbf{r}_N}^* \nonumber\\
&\times\bra{\tilde{\mathbf{r}}_1,\ldots,\tilde{\mathbf{r}}_N} U(t_2-t_1) \ket{\overline{\mathbf{r}}_1,\ldots,\overline{\mathbf{r}}_N}\,.
\end{align}
Here, $d^3r_{1\ldots N}$ is a short notation for $d^3r_1 \ldots d^3 r_N$ (and analogous for $\tilde{r}$ and $\overline{r}$), and $\psi(\mathbf{r}_1,\ldots,\mathbf{r}_N,t)$ is the wave function of the many-body system. Further, the volume $\overline{V}$ in the integration over $\overline{r}_{1\dots N}$ indicates that the point at which the coordinates $\mathbf{\overline{r}}_{1\dots N}$ coincide with the coordinates $\mathbf{r}_{1\dots N}$ should be omitted, in generalization of the condition $j\neq m$ in Eq.~(\ref{eqn: CohCorr}). 

The interpretation of Eqs.~(\ref{eqn: ProjDCF}, \ref{eqn: CohDCF}) is straightforward in comparison with Eqs.~(\ref{eqn: ProjCorr},\ref{eqn: CohCorr}), if one notes the correspondences 
\begin{subequations}
\begin{align}
a_j  &\leftrightarrow \rho(\mathbf{r}';\mathbf{r}_1,\ldots,\mathbf{r}_N) \,,\\
b_l  &\leftrightarrow  \rho(\mathbf{r}'+\mathbf{r};\tilde{\mathbf{r}}_1,\ldots,\tilde{\mathbf{r}}_N) \,,\\
c_j(t_1) &\leftrightarrow \psi(\mathbf{r}_1,\ldots,\mathbf{r}_N,t_1) \,,\\
c_m(t_1) &\leftrightarrow \psi(\overline{\mathbf{r}}_1,\ldots,\overline{\mathbf{r}}_N,t_1) \,,\\
|a_j\rangle &\leftrightarrow \ket{\mathbf{r}_1,\ldots,\mathbf{r}_N}\,,\\
|b_l\rangle &\leftrightarrow \ket{\tilde{\mathbf{r}}_1,\ldots,\tilde{\mathbf{r}}_N} \,,\\
|a_m\rangle &\leftrightarrow \ket{\overline{\mathbf{r}}_1,\ldots,\overline{\mathbf{r}}_N}\,.
\end{align}
\end{subequations}
The integrals instead of the sums arise from the continuous nature of the eigenvalue spectrum. 

From Eq.~(\ref{eqn: ProjDCF}) we find that the projective part $\mathscr{G}(\mathbf{r},t_1,t_2)$ of the DCF is determined by the probability of finding the particles in an initial configuration $\{ \mathbf{r}_1,\ldots,\mathbf{r}_N \}$ at $t_1$, given by the modulus square of the many-body wavefunction of the system, times the probability that the particles moved from this configuration at $t_1$ to the configuration $\{ \tilde{\mathbf{r}}_1,\ldots,\tilde{\mathbf{r}}_N \}$ at $t_2$, given by the modulus square of the transition matrix element between the initial and final configurations. This product of probabilities is just the joint probability of finding the particles in the two above mentioned configurations at different times.

The coherent part $\Gamma(\mathbf{r},t_1,t_2)$ in Eq.~(\ref{eqn: CohDCF}) instead does not depend on the probabilities, but on the probability amplitudes that the particles go from either of two initial configurations, $\{ \mathbf{r}_1,\ldots,\mathbf{r}_N \}$ and $\{\overline{\mathbf{r}}_1,\ldots,\overline{\mathbf{r}}_N \}$ at $t_1$ to a final configuration $\{ \tilde{\mathbf{r}}_1,\ldots,\tilde{\mathbf{r}}_N \}$ at $t_2$, and on the coherence of the many-body wavefunction at $t_1$ between the two different initial configurations of the particles.

After having established the splitting of the DCF into the projective and the coherent part in Eq.~(\ref{eq: splitG}), we analogously split the corresponding ISF in Eq.~(\ref{eqn: DefISF}) into two parts,
\begin{align}
\label{eq:ssplit}
 S(\mathbf{p},t_1, t_2) &= S_{\mathscr{G}}(\mathbf{p},t_1,t_2) + S_{\Gamma}(\mathbf{p},t_1,t_2)\,,
\end{align}
where $S_{\mathscr{G}}$ and $S_{\Gamma}$ are the parts of the ISF arising from the projective contribution $\mathscr{G}$ and the coherent contribution $\Gamma$ to $G$.

\subsection{\label{sec:backaction-tdi}Measurement backaction in TDI}
In our previous work~\cite{PhysRevLett.122.025301}, we showed that the DCF can have a non-vanishing imaginary part, indicating the presence of quantum correlations in the target. In the following, we discuss this issue further related to the projective and the coherent part of the DCF and their contributions to the ISF.

To start the discussion, we recall that for a general quantum mechanical target, the DCF and the ISF have the following properties~\cite{VANHOVE,PhysRevLett.122.025301}
\begin{subequations}
\begin{align}
 G(\mathbf{r},t_1,t_2)^* &= G(-\mathbf{r},t_2,t_1)\,,\\
 S(\mathbf{p},t_1,t_2)^* &= S(\mathbf{p},t_2,t_1)\,.
\end{align}
\end{subequations}
For a classical target, the potentially non-commuting density operators in the DCF are replaced by their classical counterparts, such that the DCF becomes real, and the DCF and ISF have additional symmetries
\begin{subequations}
 \begin{align}
 G_{cl}(\mathbf{r},t_1,t_2)^* &= G_{cl}(-\mathbf{r},t_2,t_1) = G_{cl}(\mathbf{r},t_1,t_2)\,,\\
 S_{cl}(\mathbf{p},t_1,t_2)^* &= S_{cl}(\mathbf{p},t_2,t_1)= S_{cl}(-\mathbf{p},t_1,t_2)\,.
\end{align}
\end{subequations}
From Eqs.~(\ref{eqn: ProjDCF}), (\ref{eqn: CohDCF}) and (\ref{eq:ssplit}) we find that the projective parts $\mathscr{G}$ and its contribition $S_{\mathscr{G}}$ to the ISF obey the classical symmetries, whereas the coherent parts $\Gamma$ and the corresponding ISF part $S_{\Gamma}$ in general only follow the more restrictive quantum mechanical symmetries.
As shown in~\cite{PhysRevLett.122.025301}, these symmetries can be exploited to test for a quantum mechanical nature of the target, by comparing measurements with opposite transfer momenta $\pm \mathbf{p}$. This clearly demonstrates that QTDI is not restricted to the measurement of the projective part of ${\mathscr{G}}$ of G and the corresponding contribution $S_{\mathscr{G}}$ to the ISF.

Going one step further, the phase control in QTDI offers the possibility to reconstruct the ISF for given $\mathbf{p},t_1$ and $t_2$ completely, as discussed in Sec.~\ref{sec:recovery}. This allows one in principle to access experimentally the imaginary part of the DCF, and not only to check for its presence via the above-mentioned symmetry conditions. Indeed, by defining the symmetric and anti-symmetric parts of the ISF
\begin{equation}
S^{\pm}(\mathbf{p},t_1,t_2)=\frac{S(\mathbf{p},t_1,t_2) \pm S(-\mathbf{p},t_1,t_2)}{2}\,,
\end{equation}
we can express for the Fourier transform of  $\text{Im}[\Gamma(\mathbf{r},t_1,t_2)]$ as 
\begin{align}
\label{eqn: FourImG}
\int d^3 r \: &\text{Im}[\Gamma(\mathbf{r},t_1,t_2)] \: e^{i\mathbf{p}\cdot \mathbf{r}}  \nonumber\\
&=\text{Im}[S^+ (\mathbf{p},t_1,t_2)]+\text{Re}[S^- (\mathbf{p},t_1,t_2)]\,.
\end{align}
Therefore, if a sufficiently large region of the exchanged momentum space can be probed in an experiment, the $\text{Im}[\Gamma(\mathbf{r},t_1,t_2)]$ can be recovered from the data by an inverse Fourier transform of Eq.~(\ref{eqn: FourImG}).
As a result, we find that QTDI provides a way to measure the complete complex-valued DCF without backaction.

\begin{figure}[t]
  \centering
   \includegraphics[]{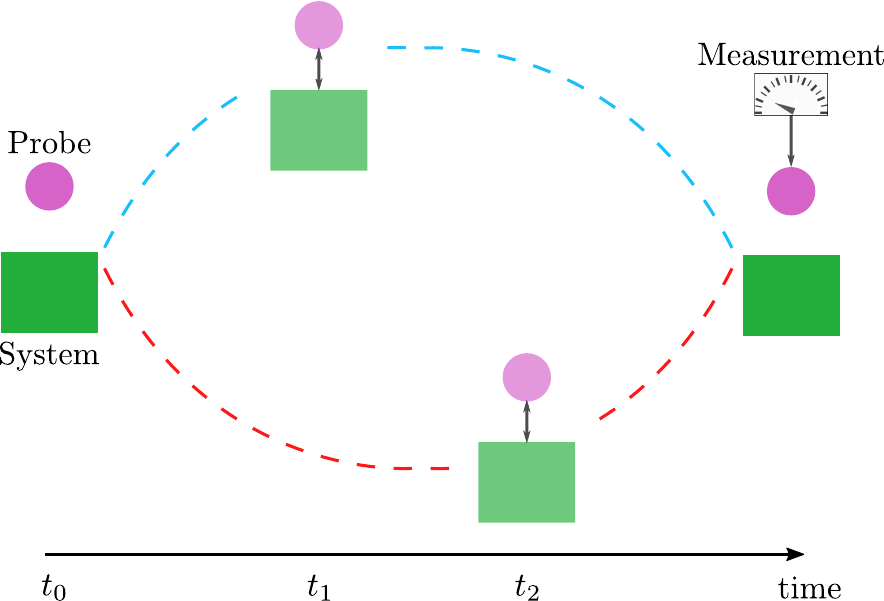}
	\caption{(Color online) Schematic representation of the branching dynamics. The system and the probe are initially uncoupled, the following evolution of the joint system branches out into two possible histories: either the probe and the system couple at time $t_1$ (blue branch) or they couple at time $t_2>t_1$ (red branch). At a later time a measurement on the probe that is blind to the coupling time, recomposes the information gathered on the system along the two possible histories.}
\label{fig: scheme}
\end{figure}

\section{Conclusions and Outlook}
In this work we have extended the analysis given in \cite{PhysRevLett.122.025301} of the working of the QTDI technique for the measurement of space-time correlations among the particles of a quantum system. As discussed in the introduction, the measurement of two-time correlations between two generic observables of a quantum system in general cannot be accomplished by probing the observables of interest at two consecutive times, because the first probing would back-act on the system altering its successive dynamics and therefore also the results of the second probing. As a result, the quantity obtained by correlating the information obtained in consecutive measurements would not represent faithfully the correlations that develops spontaneously in the system.

Throughout our analysis, we have explicitly demonstrated this for the case in which the probings consist of direct projective measurements. In Sec.~\ref{sec: Quantum Dynamical Correlation Functions} we have put forward an expression for generic two-time correlations as a sum of two terms that we named projective and coherent parts. Consecutive direct projective measurements would give access only to the former part, the second being destroyed by the collapse of the state of the system due to the earlier measurement. We also found that the coherent part endows the time correlation function with an imaginary part. This is a distinctive feature of quantum-mechanical time-correlation functions, in direct contrast to their classical counterparts, and is completely lost under direct projective measurements. If applied to QTDI,  the experimentally accessible ISF reflects contributions from the imaginary part of the DCF, the latter spoiling the symmetry properties of the former.

On the other hand, it is anyways necessary to interact with the system in order to extract information on the observables, so that back-action is inevitable. Despite of that, our analysis of TDI shows that the full ISF can be retrieved from measurement, that is the one accounting for both the projective and coherent part of the DCF. How is this possible? It turns out that the dynamics of the target and the incoming radiation branches out in two possible pathways, see Fig.~\ref{fig: scheme}. In each repetition of the experiment, either the target is probed by the earlier incoming wavepacket or by the later one, but not consecutively by both. As a result, the second scattering is not modified by a potential backaction of the first scattering, and therefore it retrieves information about the distribution of particles as if the first back-action did not happen. Afterwards, the two pathways are recombined in such a way that which-way information is lost. As a result, the  scattered intensity  acquires a dependence on the distribution of particles at the different scattering times in the form of the sought-after ISF.

For the case of a single incident x-ray photon, this branching can be interpreted as the formation of an intermediate time-bin entangled state~\cite{PhysRevLett.88.070402,PhysRevLett.103.017401}, in which the single photon is in a superposition of the two pulses interacting with the target at different times.  However, it presently is not possible to generate such single x-ray photons with sufficient count rates for applications in QTDI. For this reason, we have extended QTDI to the case of weak classical continuous-mode coherent states, as an approximation of the classical fields delivered by pulsed accelerator-based x-ray sources. Interestingly, we could demonstrate that also in this case, QTDI provides access to the full quantum mechanical ISF, because the requirement of the single incident x-ray photon can be replaced in a single-interaction approximation appropriate for the case of weak incident fields.

Having established the capability of TDI to faithfully probe the ISF of quantum mechanical targets, we finally propose an extension of the scheme which makes it possible to extract this quantity directly from data. This would allow the reconstruction of the imaginary part of the DCF, obtaining an experimental signature of the quantumness of correlations among particles. A caveat is necessary here: in \cite{PhysRevLett.122.025301} we studied a concrete quantum mechanical model system exhibiting states that give rise to a purely real DCF. This example shows that in case a vanishing imaginary part of the DCF is found from experimental data, the classicality of the system cannot be safely claimed unless quantum effects in the correlations among particles can be excluded on a firm physical basis. Therefore comparison between detailed theoretical models and experimental data is necessary, and our proposed modification of TDI can be of great help for this purpose.

From a general perspective, even though our analysis focused on the specific case of QTDI, its mechanism to access the full quantum-mechanical correlations without backaction is general and could inspire new methods for measuring other correlations. The key idea is to couple the system under investigation to an ancilla system in such a way that the dynamics of the whole can unfold along  two alternative pathways, in each of which the ancilla probes the observable of interest at a different time, see Figure~\ref{fig: scheme}. The expectation value of measurements on the ancilla that do not discriminate the time of interaction will then depend on the observable of the system evaluated at both times in the form of the sought-for correlation.

\begin{acknowledgments}
This work is part of and supported by the DFG Collaborative Research Centre ``SFB 1225 (ISOQUANT).''
\end{acknowledgments}

\appendix

\section{\label{app-perturb}Perturbative calculation}
In this appendix we show the explicit first-order perturbative calculations leading to $\ket{\delta \Psi_\alpha (t)}$ and $\ket{\delta \Psi_\beta (t)}$ in Eqs.~(\ref{eqn: ExplDeltaAleph}) and (\ref{eqn: ExplDeltaGimel}).

The first-order correction to the evolution of the initial state (\ref{eqn: InitState}) explicitly reads
\begin{align}
\label{eqn: AppFirstOrdImpl}
&\ket{\Psi^{(1)}(t)}=
-\frac{i}{\hbar} \int_{0} ^t dt'  H_I (t) \ket{\psi}\ket{\alpha,\beta} \nonumber\\
&= -i\frac{r_e c^2}{4\pi} \int_0 ^t dt' \int d^3k \int d^3k' \int_V d^3 r \frac{1}{\sqrt{\omega_k \omega_{k'}}}  \nonumber \\
& \times \rho(\mathbf{r},t')\ket{\psi} \: e^{-i(\mathbf{k}\cdot \mathbf{r}-\omega_k t')} e^{i(\mathbf{k}'\cdot \mathbf{r}-\omega_{k'} t')}a^\dagger _\mathbf{k} a_{\mathbf{k}'}\ket{\alpha,\beta}\,.
\end{align}
The action of the destruction operator $a_{\mathbf{k}'}$ on the inital coherent state is calcualted by commuting the latter with the displacement operators defining  the coherent state, that is
\begin{align}
\label{eqn: ActonCohState}
a_{\mathbf{k}'}\ket{\alpha,\beta}&=a_{\mathbf{k}'}D(\alpha)D(\beta)\ket{0} \nonumber\\
&= \bigg( D(\alpha)D(\beta)a_{\mathbf{k}'}+D(\alpha)[a_{\mathbf{k}'},D(\beta)] \nonumber\\
&\qquad +[a_{\mathbf{k}'},D(\alpha)]D(\beta)\bigg) \ket{0}\,.
\end{align}
It can be easily checked that the commutators in the last line are non-zero only for $\mathbf{k}' \parallel z$, for which they give
\begin{align}
\label{eqn: CommDispl}
[a_{\mathbf{k}'},D(\alpha)]&=\tilde{\alpha}(k'_z)D(\alpha) \,,\nonumber \\
[a_{\mathbf{k}'},D(\beta)]&= \tilde{\beta}(k'_z)D(\beta)\,.
\end{align}
Substitution (\ref{eqn: CommDispl}) into (\ref{eqn: ActonCohState}) leads to
\begin{equation}
\label{eqn: AppDesOnCoh}
a_{\mathbf{k}'}\ket{\alpha,\beta}=\big( \tilde{\alpha}(k'_z) + \tilde{\beta}(k'_z)\big) \ket{\alpha,\beta}\,.
\end{equation}
so that (\ref{eqn: AppFirstOrdImpl}) splits into the following sum
\begin{align}
\label{eqn: AppFirstOrd1}
\ket{\Psi^{(1)}(t)}&=-i\frac{r_e c^2}{4\pi \sqrt{\mathscr{A}}} \int_0 ^t dt' \int_V d^3 r  \rho(\mathbf{r},t')\ket{\psi} \nonumber \\
& \times  \int dk'_z \frac{e^{i(k'_z z-\omega_{k'} t')}}{\sqrt{\omega_{k'}}} \big[ \tilde{\alpha}(k'_z) + \tilde{\beta}(k'_z) \big] \nonumber \\
&\times\int d^3k  \frac{e^{-i(\mathbf{k}\cdot \mathbf{r}-\omega_k t')}}{\sqrt{\omega_k}}  a^\dagger _\mathbf{k} \ket{\alpha,\beta} \,.
\end{align}
The two terms at the right-hand side of the last equality will respectively give rise to $\ket{\delta \Psi_\alpha}$ and $\ket{\delta \Psi_\beta}$ given in the main text. As the calculations proceed in the same way for both these terms, in the following we will use the letter $l$ to refer to either of the $\alpha$ and $\beta$ wavepackets.

Due to the peaked spectral shape of the function $\tilde{l}(k'_z)$, the frequency factor $\omega_{k'}^{-\frac12}$ can be approximated as the constant $\omega_0 ^{-\frac12}$. Moreover the integral over $k'_z$ reduces to the spatio-temporal shape of the $l$ wavepacket
\begin{equation}
\label{eqn: SpaceTimeWavepack}
\int_{-\infty} ^\infty dk \, \tilde{\alpha}(k) e^{i (kz-\omega_k t)} = \alpha(z-ct)e^{i (k_0 z - \omega_0 t + \phi_\alpha)}\,,
\end{equation}
such that the two expressions in Eq.~(\ref{eqn: AppFirstOrd1}) become
\begin{align}
\label{eqn: AppFirstOrd2}
-i&\frac{r_e c^2}{4\pi \sqrt{\omega_0 \mathscr{A}}} \int_0 ^t dt' \int_V d^3 r \: \rho(\mathbf{r},t')\:\ket{\psi}
\: l(z-ct') \nonumber\\
&\times e^{i(k_0 z-\omega_0 t'+\phi_l)} \int d^3k  \:\frac{e^{-i(\mathbf{k}\cdot \mathbf{r}-\omega_k t')}}{\sqrt{\omega_k}}  \:a^\dagger _\mathbf{k} \:\ket{\alpha,\beta}\,.
\end{align}
Now, the product $\rho(\mathbf{r},t')\alpha(z-ct')$ is non-zero only in the time interval needed for the envelope of the wavepacket to cross the target along the incidence direction $z$, that is as soon as the support of the moving envelope $l(z-ct')$ is contained in the volume of the target.\\
By assumption, the dynamics of the target can be considered frozen during this crossing time, so that the particle density operator can be taken as constant to its value at the time $t_l$ at which the wavepacket reaches the edge of the target. As a consequence, the density operator can be brought out of the time integral and Eq.~(\ref{eqn: AppFirstOrd2}) becomes
\begin{align}
\label{eqn: AppFirstOrd3}
-i&\frac{r_e c^2}{4\pi \sqrt{\omega_0 \mathscr{A}}} \int_V d^3 r  \rho(\mathbf{r},t_l)\ket{\psi} \int d^3k \frac{e^{-i\mathbf{k}\cdot \mathbf{r}}}{{\sqrt{\omega_k}} }  \nonumber \\
& \times \int_0 ^t dt' 
l(z-ct')e^{i(k_0 z-\omega_0 t'+\phi_l)}  e^{i\omega_k t'} a^\dagger _\mathbf{k} \ket{\alpha,\beta}\,.
\end{align}
Extending the time integration to the infinity, which can be done with no appreciable error as the original time boundaries are such to grant that the wavepacket is either well downstream or upstream of the target, the time integral in the last line gives
\begin{align}
\label{eqn: AppTimeInteg}
\int_{-\infty} ^{+\infty} dt' 
&l(z-ct')e^{i(k_0 z-\omega_0 t'+\phi_l)}  e^{i\omega_k t'}\nonumber\\
&=\frac{e^{i |\mathbf{k}| z }}{c} \tilde{l}(|\mathbf{k}|)\,,
\end{align}
where we have used the dispersion relation $\omega_k=c|\mathbf{k}|$.
Substituting Eq.~(\ref{eqn: AppTimeInteg}) into Eq.~(\ref{eqn: AppFirstOrd3}), and taking into account again the finite width of $\tilde{l}$, we finally obtain Eqs.~(\ref{eqn: ExplDeltaAleph}) and (\ref{eqn: ExplDeltaGimel}).

\section{\label{app-signal}Derivation of the detection amplitude}
In this Appendix, we derive the expression Eq.~(\ref{eqn: DetectRateFinal}) for the expectation value of the intensity at the detector. We again consider the notation $l$ (=$\alpha,\beta$) for the two wavepackets. The positive frequency part of the electric field is given by
\begin{equation}
\label{eqn: PosEField}
E^{(+)}(\mathbf{R},t)\equiv -i \sqrt{\frac{\hbar}{2(2\pi)^3 \epsilon_0}}\int d^3 q\,\sqrt{\omega_q} a_\mathbf{q} e^{i(\mathbf{q}\cdot \mathbf{R}-\omega_k t)}\,.
\end{equation}
Therefore, the detection amplitude reads
\begin{align}
\label{eqn: DetAmplFull}
&\bra{0}E^{(+)}(\mathbf{R},t)\ket{\delta \Psi_l (t)} =\sqrt{\frac{\hbar}{2(2\pi)^3 \epsilon_0 \mathscr{A}}} \frac{r_e c}{4\pi\omega_0}\nonumber \\
& \times \int_V d^3r \: \rho(\mathbf{r},t_l) \:\ket{\psi} \:\int d^3 q \int d^3k\nonumber \\
&\times \sqrt{\omega_q}\tilde{l}(|\mathbf{k}|)e^{i(\mathbf{q}\cdot \mathbf{R}-\omega_k t)} e^{-i(\mathbf{k}\cdot\mathbf{r}-\omega_k \frac{z}{c})}\bra{0}a_\mathbf{q}a^\dagger _{\mathbf{k}} \ket{0}\,.
\end{align}
The expectation value on the vacuum state appearing on the right-hand side of (\ref{eqn: DetAmplFull}) is
\begin{equation}
\bra{0}a_\mathbf{q}a^\dagger _{\mathbf{k}} \ket{0}=\delta({\mathbf{q}-\mathbf{k}})\,,
\end{equation}
such that we are left with one integration over the photon momentum
\begin{equation}
\label{eqn: MomentumInt}
\int d^3 k \sqrt{\omega_k}\tilde{l}(|\mathbf{k}|)e^{i\mathbf{k}\cdot( \mathbf{R}-\mathbf{r})} e^{-i(|\mathbf{k}|(ct- \frac{z}{c})}\,.
\end{equation}
Using spherical coordinates and again approximating powers of $\omega_k$ or $|\mathbf{k}|$ by their respective peak values  in the spectrally narrow wavepacket, the integral (\ref{eqn: MomentumInt}) can be easily calculated. For the angular part, we obtain
\begin{align}
\label{eqn: Integ1}
&\sqrt{\omega_0}\int_0 ^\infty d |\mathbf{k}|\:|\mathbf{k}|^2 \:\tilde{l}(|\mathbf{k}|) \:e^{-i|\mathbf{k}|(ct-z)} 
\nonumber \\
& \qquad \times\int_0 ^\pi d\theta \int_0 ^{2\pi} d\phi \sin\theta\: e^{i|\mathbf{k}||\mathbf{R}-\mathbf{r}|\cos\theta} \nonumber \\
=&2\pi\sqrt{\omega_0}\int_0 ^\infty d |\mathbf{k}|\,|\mathbf{k}|\: \tilde{l}(|\mathbf{k}|) \:e^{-i|\mathbf{k}|(ct- z)} \nonumber \\
& \qquad \times \bigg( \frac{e^{i|\mathbf{k}||\mathbf{R}-\mathbf{r}|}-e^{-i|\mathbf{k}||\mathbf{R}-\mathbf{r}|}}{i|\mathbf{R}-\mathbf{r}|} \bigg)\,.
\end{align}
The negative exponential in the round brackets is dropped hereafter because it would correspond to an ingoing spherical wave, that is to a photon traveling from the detector to the target. Performing the remaining integration over the magnitude of the wave vector,   Eq.~(\ref{eqn: MomentumInt}) becomes
\begin{align}
\label{eqn: Integ2}
&\frac{2\pi}{ic|\mathbf{R}-\mathbf{r}|}\sqrt{\omega^3_0}\int_0 ^\infty d |\mathbf{k}|\, \tilde{l}(|\mathbf{k}|) e^{i|\mathbf{k}|(|\mathbf{R}-\mathbf{r}|+z -ct )} \nonumber \\
=&
\frac{2\pi}{ic}\sqrt{\omega^3_0} \, l\big( |\mathbf{R}-\mathbf{r}|+z -ct  \big) \, \frac{e^{i |\mathbf{k}_0||\mathbf{R}-\mathbf{r}|}}{|\mathbf{R}-\mathbf{r}|}e^{i|\mathbf{k}_0|z}e^{i|\mathbf{k}_0|ct} \nonumber\\
\simeq&\frac{2\pi}{ic}\sqrt{\omega^3_0} \, l\big( |\mathbf{R}| -ct  \big) \frac{e^{i|\mathbf{k}_0||\mathbf{R}|}}{|\mathbf{R}|}e^{-i|\mathbf{k}_0|\hat{\mathbf{R}}\cdot\mathbf{r}}e^{i|\mathbf{k}_0|z}e^{i|\mathbf{k}_0|ct}\,,
\end{align}
where the last line has been obtained using the assumption that the detector is placed far away from the target, that is $|\mathbf{R}| \gg |\mathbf{r}|$.

Defining the exchanged momentum as $\mathbf{p}=k_0(\hat{\mathbf{R}}-\hat{\mathbf{z}})$, and substituting Eq.~(\ref{eqn: Integ2}) into (\ref{eqn: DetAmplFull}), we obtain the final form of the detection amplitudes Eq.~(\ref{eqn: Integ2}) given in section~(\ref{sec: The recorded intensity})
\begin{align}
\label{eqn: App-DetAmpl}
&\bra{0}E^{(+)}(\mathbf{R},t)\ket{\delta \Psi_l (t)}  \nonumber \\
&\quad=\frac{r_e}{2}\sqrt{\frac{\hbar \omega_0}{2(2\pi)^3 \epsilon_0 \mathscr{A}}}\frac{ e^{i(k_0 R -\omega_0 t)}}{R}\: e^{i\phi_l}\,l(R-ct) \nonumber \\
&\qquad \times \int_V d^3r \, \rho(\mathbf{r},t_l)\ket{\psi}e^{-i \mathbf{p}\cdot\mathbf{r}}\,.
\end{align}

\section{\label{app-equiv}Detection signal without substitution}
In this Appendix, we derive the detected intensity Eq.~(\ref{eqn: DetectRateFinal}) without the substitution of $a^\dagger _{\mathbf{k}}\ket{\alpha,\beta}$ by $a^\dagger _{\mathbf{k}}\ket{0}$  which was used in Eq.~(\ref{eqn: ScattRateAmpl}) to simplify the analysis. We find that the result obtained in both ways for the expected value of the intensity are the same, showing the validity of taking the field part of the final state as a superposition of single-photon spherical wavepackets despite the many-photon content of the incoming continuous-mode coherent state.
Without applying the substitution, the intensity would be calculated as
\begin{align}
\label{eqn: IntensityFull}
&\sum_{h,l=\alpha,\beta} \bra{\delta \Psi_h (t)}E^{(-)}(\mathbf{R},t)E^{(+)}(\mathbf{R},t)\ket{\delta \Psi_l (t)} \nonumber \\
&=\sum_{h,l=\alpha,\beta} \bigg( \frac{r_e c}{4 \pi \omega_0 \sqrt{\mathscr{A}}} \bigg)^2 \nonumber \\
&\times \int_V d^3 r' \bra{\psi} \rho(\mathbf{r}',t_h)
\int d^3 k' \, h^*(|\mathbf{k}|) \, e^{i \mathbf{k}'\cdot \mathbf{r}} e^{i \omega_{k'}(t - \frac{z'}{c})} \nonumber \\
&\times \int_V d^3 r   \rho(\mathbf{r},t_l) \ket{\psi}
\int d^3 k \, l(|\mathbf{k}|)  \, e^{-i \mathbf{k} \cdot \mathbf{r}} e^{-i \omega_{k}(t - \frac{z}{c})} \nonumber\\
&\times \bra{\alpha,\beta}a_{\mathbf{k}'} E^{(-)}(\mathbf{R},t)E^{(+)}(\mathbf{R},t)a^\dagger _\mathbf{k}\ket{\alpha,\beta}\,.
\end{align}
The continuous-mode coherent state $\ket{\alpha,\beta}$ is an eigenstate of the positive frequency part of the electric field, the eigenvalue being the value of the field at the position of the detector at time $t$. As the detector is assumed to be out of reach of the incoming pulse, this value can be taken to be zero so that
\begin{equation}
\label{eqn: NoField}
E^{(+)}(\mathbf{R},t)\ket{\alpha,\beta}=0\,.
\end{equation}
Therefore, in order to evaluate the inner product in Eq.~(\ref{eqn: IntensityFull}) we move the negative frequency part of the field to the left and the positive frequency part to the right, by commuting these operators with the creation and destruction operators. By doing so we find that
\begin{align}
\label{eqn: ExpValCoh}
&\bra{\alpha,\beta}a_\mathbf{k} E^{(-)}(\mathbf{R},t) E^{(+)}(\mathbf{R},t)a^\dagger _\mathbf{k'}\ket{\alpha,\beta} \nonumber \\
&= \bra{\alpha,\beta} \big( E^{(-)}(\mathbf{R},t)a_\mathbf{k}+[a_\mathbf{k},E^{(-)}(\mathbf{R},t)] \big) \nonumber \\
& \qquad \times \big( a^\dagger _\mathbf{k'} E^{(+)}(\mathbf{R},t)  + [E^{(+)}(\mathbf{R},t),a^\dagger _\mathbf{k'}] \big) \ket{\alpha,\beta}
\nonumber \\
&= \frac{\hbar}{2 (2\pi)^3 \epsilon_0 }\sqrt{\omega_k \omega_{k'}}e^{-i (\mathbf{k}\cdot \mathbf{R}-\omega_k t)}e^{i (\mathbf{k}' \cdot \mathbf{R}-\omega_{k'} t)}\,,
\end{align}
where Eq.~(\ref{eqn: NoField}) has been used and the commutators have been calculated starting from Eq.~(\ref{eqn: PosEField}). Substituting Eq.~(\ref{eqn: ExpValCoh}) into Eq.~(\ref{eqn: IntensityFull}) we find 
\begin{align}
\label{eqn: FullCalcInt}
&\sum_{l,h=\alpha,\beta} \frac{\hbar}{2 (2\pi)^3 \epsilon_0 \mathscr{A}} \bigg( \frac{r_e c}{4 \pi \omega_0} \bigg)^2  \nonumber \\
& \times \int_V d^3 r' \bra{\psi} \rho(\mathbf{r}',t_h)
\int d^3 k' \, \sqrt{\omega_{k'}} \tilde{h}^*(|\mathbf{k}'|) \, e^{-i \mathbf{k}'\cdot (\mathbf{R}-\mathbf{r}')} \nonumber\\
& \; e^{i \omega_{k'}(t - \frac{z'}{c})} \nonumber \\
& \times \int_V d^3 r   \rho(\mathbf{r},t_l) \ket{\psi} \int d^3 k \, \sqrt{\omega_{k}} \tilde{l}(|\mathbf{k}|)  \, e^{i \mathbf{k} \cdot (\mathbf{R}-\mathbf{r})} e^{-i \omega_{k}(t - \frac{z}{c})}\,.
\end{align}
By computing the integrals over the photon momentum as done in Eqs.~(\ref{eqn: Integ1})-(\ref{eqn: Integ2}), Eq.~(\ref{eqn: FullCalcInt}) reduces to
\begin{align}
\label{eqn: FullCalcInt1}
&\sum_{l,h=\alpha,\beta}  \frac{\hbar}{2 (2\pi)^3 \epsilon_0 \mathscr{A}} \bigg( \frac{r_e c}{4 \pi \omega_0} \bigg)^2 \nonumber  \\
& \times \frac{ e^{-i(k_0 R -\omega_0 t)}}{R}\: e^{-i\phi_h}\,h(R-ct)^*
\int_V d^3r' \, \bra{\psi} \rho(\mathbf{r}',t_h)e^{i \mathbf{p}\cdot\mathbf{r}'} \nonumber \\
& \times \frac{ e^{i(k_0 R -\omega_0 t)}}{R}\: e^{i\phi_l}\,l(R-ct)
\int_V d^3r \,  \rho(\mathbf{r},t_l)\ket{\psi} e^{-i \mathbf{p}\cdot\mathbf{r}} \,.
\end{align}
We notice that the two factors making up the generic term of the sum above, are the same as the RHS of Eq.~(\ref{eqn: App-DetAmpl}). Therefore, unfolding the sum one obtains the very same expression Eq.~(\ref{eqn: DetectRateFinal}) given in the main text for the intensity of the field, which has been originally calculated using the single scattered photon approximation.

\bibliographystyle{myprsty}
\bibliography{Bibliography}

\end{document}